\documentstyle[12pt]{article}
\newcommand{\be}{\begin{equation}}
\newcommand{\en}{\end{equation}}
\newcommand{\bea}{\begin{eqnarray}}
\newcommand{\ena}{\end{eqnarray}}

\newcommand{\hbo}{\hbox to 1 true cm {\hfill } }

\newcommand{\Tr}{\hbox{Tr}}

\def\dslash{\partial\kern-.5em\slash}
\def\kslash{k\kern-.5em\slash}

\begin{document}
\vglue 1truecm

\vbox{
UNIT\"U-THEP-15/1993  \hfill October 19, 1993.
}

\vfil
\centerline{\bf \large Natural Slow-Roll Inflation$^1$}

\bigskip
\centerline{ R.\ F.\ Langbein, K.\ Langfeld, H.\ Reinhardt,
   L.\ v.\ Smekal }
\medskip
\centerline{Institut f\"ur theoretische Physik, Universit\"at
   T\"ubingen}
\centerline{D--74076 T\"ubingen, Germany}
\bigskip

\vfil
\begin{abstract}

It is shown that the non-perturbative dynamics of a phase change to
the non-trivial phase of $\lambda\varphi^4$-theory in the early
universe can give rise to slow-rollover inflation without recourse to
unnaturally small couplings.

\end{abstract}

\vfil
\hrule width 5truecm
\vskip .2truecm
\begin{quote}
$^1$ Supported by DFG under contract Re $856/1 \, - \, 1$
\end{quote}
\eject
{\it 1.\ Introduction \/ }

The inflation paradigm~\cite{gu80,as82,li83,pi84,hol84} solves a
number of cosmological problems, including the horizon problem, the
flatness-age problem and the problem of the origin of density
inhomogeneities~\cite{ko90}.  Its essential feature is a super-luminal
expansion of the universe at very early times.  Models of inflation
generally rely on the dynamics of some scalar field (the inflaton),
which can also be thought of as an order parameter for the possible
phases of a scalar field. At early times it is assumed that the
inflaton is displaced from the absolute minimum of its potential, and
the potential energy density thus produced is assumed to dominate.
Such a vacuum energy density, inserted into the (standard Einstein)
gravitational action, creates a period of exponential expansion of the
universe with time until the scalar field attains its absolute
minimum. One of the perhaps more natural mechanisms for localising the
field away from its absolute minimum is based on the
temperature-dependent effective potential (of a form similar to that
in figure~1) of the scalar field, where at high temperature the
minimum of the potential is at the origin but at temperatures below a
critical temperature the absolute minimum takes some other value (any
remaining local minimum at the origin being labeled the false vacuum).
The first inflation models assumed a strong first order phase
transition from the false to true vacuum~\cite{gu80} and the time
before this tunneling process occurred gives the duration of the
exponential expansion.  However, if this tunneling time is too large,
the phase transition, turning false vacuum into true vacuum, will
never be completed. In fact, a detailed calculation~\cite{gu83} shows
that the inflation model with a strong first order phase transition is
very unlikely.  Subsequent to this, many models were designed to
overcome this problem whilst preserving the essential ingredients of
inflation~\cite{pi84,hol84}. One of the more successful approaches
relies on a mechanism originally proposed by Albrecht and Steinhardt,
and Linde where the effective potential of the order parameter has an
extremely flat region near the false vacuum minimum implying that the
evolution of the inflaton to the true vacuum is via a process of
``slow rolling''~\cite{as82,li83}. Such models successfully implement
inflation but require unnaturally small couplings ($\sim10^{-12}$) to
satisfy the bounds on density fluctuations imposed by the cosmic
microwave background radiation (CMBR). The mechanism arising from
natural particle physics models which could produce such small
couplings, or more generally, an extremely flat part of the potential,
is not yet fully understood, and will be the subject of this letter.

Recently, some insight into the vacuum structure of massless $\varphi
^{4}$-theory has been gained~\cite{la1,la2,la3}. The theory has two
phases, a trivial phase and a non-trivial phase with non-vanishing
condensate. The effective potential of the scalar field $\varphi $ was
calculated in a renormalisation group invariant approach in~\cite{la1}
and it was found that in the trivial phase it coincides with the
one-loop renormalisation group improved effective potential. However,
in the non-perturbative phase, a dynamical mass for the scalar field is
generated, and the vacuum energy density agrees with the predictions
from the scale anomaly.  The effective potential is convex~\cite{la1}
in both phases, implying $\langle \varphi \rangle =0$ and this means that
$\langle \varphi \rangle $ is not an appropriate order parameter for
describing the vacuum structure of the theory. We will see below
that an appropriate and convenient order parameter is $\sqrt{\langle
\varphi^{2} \rangle }$. The effective potential of the scalar condensate
${\langle\varphi^{2} \rangle }$ was calculated in~\cite{la2} where it was
found that at large temperature the system is in the perturbative
phase~(see figure~1) and that at a critical temperature the energy
density of the non-perturbative phase becomes lower than that of the
trivial phase, and a phase transition to the non-trivial vacuum
occurs.

In this letter we present the effective action for the order parameter
$\sqrt{ \langle \varphi^{2} \rangle }$ at finite temperature $T$ in a
derivative expansion. We will find that the kinetic energy term
acquires a temperature-dependent dynamical prefactor, which becomes
large if $\sqrt{ \langle \varphi^{2} \rangle } < T$. This prefactor
implies that tunneling from the false vacuum state ($\sqrt{ \langle
\varphi^{2}\rangle } = 0$) to the true ground state is suppressed. The
system can be quantitatively described in terms of an auxiliary field
(with standard kinetic term) moving in a modified potential and this
modified potential is of the form used in slow-rolling-inflation
models. This can explain the slow roll phase transition of the
inflaton to the true vacuum in a natural way without relying on an
unnaturally small coupling.

\bigskip
{\it 2.\ Effective action of the scalar condensate in derivative expansion
\/ }

Starting from the Euclidean generating function for $\lambda\varphi^4$
theory
\be
Z[j]=\int{\cal D}\varphi\;\exp\{-\int d^4x\,({1\over2}\partial_\mu\varphi
\partial_\mu\varphi+{m^2\over2}\varphi^2+{\lambda\over4!}\varphi^4
-j(x)\varphi^2(x)\,)\} \label{eq:0}
\en
and then linearising the $\varphi ^{4}$-interaction with an auxiliary
field $M$ it is possible to calculate the effective potential (see
figure~1) for the scalar condensate at finite and zero
temperature~(see refs.~\cite{la1,la2,la3} for details). Here we
generalise the calculation to weakly space dependent condensates, and
evaluate the effective action in a gradient expansion. To lowest order
in the modified loop expansion~\cite{la1,la2} we have
\be
- \ln Z[j] \; = \; \int _{0}^{1/T} d \tau \; \int d^{3}x \;
\bigl( - \frac{3}{2 \lambda }
[M- m^{2} + 2 j(\vec{x}) ]^{2} \bigr) \, + \, \frac{1}{2} \Tr \ln (
-\partial ^{2} + M(\vec{x}) ) \; ,
\label{eq:1}
\en
where $T$ is the temperature in units of Boltzmann's constant and $M$
now takes its mean field value.  At finite temperature, the trace
extends over all configurations in space-time which satisfy periodic
boundary conditions in the Euclidean time direction. In Schwinger's
proper time regularisation the loop contribution is
\be
L = \frac{1}{2} \Tr \, \ln (- \partial ^{2} + M ) \; = \;
- \frac{1}{2} \Tr _{3} \; \sum _{n}
\int _{1 / \Lambda ^{2} } \frac{ds}{s} \; \exp \{
-s ( -\nabla ^{2} + M(\vec{x}) + ( 2\pi T )^{2} n^{2} \} \; ,
\label{eq:1a}
\en
where $\Lambda $ is the proper time cutoff and the trace ${\rm Tr}_3$
in (\ref{eq:1a}) extends only over the space degrees of freedom.
Using Poisson's formula we obtain the alternative form
\be
L = -\frac{1}{ 4 \sqrt{\pi } T } \int \frac{ ds }{ s^{3/2} }
\sum _{\nu } \exp ( - \frac{1}{ 4 s T^{2} } \nu ^{2} ) \, \Tr _{3}
\exp [ -s ( - \nabla ^{2} + M(\vec{x} ) ) ] \; ,
\label{eq:1b}
\en
which is more convenient for discussing the ultra-violet behaviour of
the loop~\cite{la2,la4}.  The divergences arising from the loop in
(\ref{eq:1}) can be absorbed into the bare parameters $\lambda, m, j$.
Here we generalise the renormalisation procedure of \cite{la2} to
space-time dependent sources $j(x)$, i.e.,
\bea
\frac{6}{ \lambda } \; + \; \frac{1}{16 \pi ^{2} } \,\left( \ln
\frac{\Lambda ^{2}}{\mu ^{2}} -\gamma + 1\right)  &=& \frac{6}{ \lambda _{R} }
\label{eq:2} \\
\frac{6}{\lambda } j(x) \, - \, \frac{3 m^{2} }{\lambda } \; - \;
\frac{1}{32 \pi ^{2} } \Lambda ^{2} &=& \frac{6}{ \lambda _{R} } j_{R}(x)
\, - \, \frac{ 3 m_{R}^{2} }{ \lambda _{R} }
\label{eq:3} \\
\int d^{4}x \; j^{2} (x) \; - \; m^{2} \int d^{4}x \; j(x) &=& 0 \; ,
\label{eq:4}
\ena
where $\lambda _{R}, m_{R}, j_{R}$ are the renormalised quantities. In
the following we only consider the massless case $m_{R}=0$. In this
case, equations (\ref{eq:3}) and (\ref{eq:4}) imply that
\bea
- \ln Z[j_{R}] &=& \int _{0}^{1/T} d \tau \; \int d^{3}x \;
\bigl( -\frac{3}{2 \lambda } M^{2} - \frac{6}{\lambda _{R} } j_{R} M
- \frac{1}{32 \pi ^{2} } \Lambda ^{2} M \bigr)
\label{eq:5} \\
&+& \frac{1}{2} \Tr \ln ( -\partial ^{2} + M(\vec{x}) ) \; .
\nonumber
\ena
This generating functional is finite. The quadratic and logarithmic
divergencies of the trace-term in (\ref{eq:5}) are cancelled by the
terms $- \frac{1}{32 \pi ^{2} } \Lambda ^{2} M $, $-\frac{3}{2 \lambda
} M^{2} $ respectively~\cite{la2}.  The effective action can now be
obtained by a Legendre transformation with respect to the renormalised
source $j_{R}$, i.e.,
\be
\Gamma [ \varphi ^{2}_{c}] := - \ln Z[j_{R}] + \int d^{4}x \;
\varphi ^{2}_{c}(x) j_{R}(x) \; , \hbo
\varphi ^{2}_{c}(x) :=  \frac{ \delta \, \ln Z[j_{R}] }{ \delta j_{R}(x) }
\label{eq:6}
\en
where the classical field $\varphi_c$ is defined by
\be
\varphi ^{2}_{c}(\vec{x}) \; = \; \frac{6}{\lambda _{R}} M(\vec{x}) \; .
\label{eq:7}
\en
Due to field renormalisation, $\lambda _{R} \varphi ^{2}_{c}$ is
renormalisation group invariant~\cite{la2}, and we therefore refer to
$M$ as the physical scalar condensate.  The effective action for
$M(\vec{x})$ is
\be
\Gamma [M] \; = \; \int _{0}^{1/T} d \tau \; \int d^{3}x \;
\bigl( - \frac{3}{2 \lambda } M^{2} - \frac{1}{32 \pi ^{2} } \Lambda ^{2}
M(\vec{x}) \bigr) \, + \,
\frac{1}{2} \Tr \ln ( -\partial ^{2} + M(\vec{x}) ) \; .
\label{eq:8}
\en
We wish to evaluate this effective action for weakly space-dependent
fields $M(\vec{x})$ in a gradient expansion. This can be more
systematically accomplished by using the modified heat-kernel
expansion proposed in ref. \cite{re88}. Since we are only interested
in the first two terms in the expansion however, we can follow a more
direct, though perhaps less elegant, route to the same
result. We expand $M$ around an arbitrary point $\vec{x_{0}}$, i.e.,
\be
M(\vec{x}) \; = \; M(\vec{x}_{0}) + \eta (\vec{x})
\; =: \; M_{0} + \eta (\vec{x}) \; ,
\label{eq:9}
\en
and we may assume that only derivatives of $\eta $ up to second order
contribute to the loop.  Using
\bea
e^{- K_{0} + A } & = & e^{ -K_{0} } \; + \; \int _{0}^{1} dx \;
e^{-x K_{0} } A e^{-(1-x) K_{0} }
\label{eq:10} \\
&+& \int _{0}^{1} dx \; \int _{0}^{1-x} dy \; e^{-xK_{0}} A
e^{-(1-x-y)K_{0}} A e^{-y K_{0} } \; + \; O(A^{3} ) \; ,
\nonumber
\ena
the loop (\ref{eq:1b}), up to second order in the field $\eta $, is
\bea
L & = & L_{0} \; + \; L_{1} \; + \; L_{2} \; + \; O(\eta ^{3} ),
\nonumber \\
L_{2} &=& -\frac{1}{ 4 \sqrt{\pi } T } \int \frac{ ds }{ s^{3/2} }
\sum _{\nu } \exp ( - \frac{1}{ 4 s T^{2} } \nu ^{2} ) \;
\int _{0}^{1} dx \; x \, \Tr _{3} \, [ e^{-x K_{0} } \eta
e^{-(1-x) K_{0} } \eta ] \, s^{2} ,
\label{eq:11}
\ena
where $L_{0}$ and $L_{1}$ denote the terms independent and linear
in $\eta $ and
\be
K_{0} \; = \; s \, (- \nabla ^{2} \, + \, M_{0} ) \; .
\label{eq:12}
\en
$L_{2}$ is most easily evaluated in momentum space where the
trace-term in $L_{2}$ becomes
\be
\int \frac{ d^{3}p }{ (2\pi )^{3} } \; \tilde{ \eta } (\vec{p} )
\{ \int \frac{ d^{3}k }{ (2\pi )^{3} } \;
e^{-x s ((\vec{p}+\vec{k})^{2} + M_{0} ) }
\, e^{-(1-x) s (\vec{k}^{2} + M_{0} ) } \} \, \tilde {\eta } (- \vec{p} )
\; ,
\label{eq:13}
\en
with $\tilde{\eta }$ denoting the Fourier transform of the field
$\eta (\vec{x})$. If we confine ourselves to weakly space dependent fields
$\eta $, we may expand (\ref{eq:13}) up to second order in the
external momentum $\vec{p}$. The kinetic term of the $\eta $-field
is
\be
L_{kin} \; = \; \frac{1}{ 192 \pi ^{2} T } \int ds \;
\sum _{\nu } \exp ( - \frac{1}{ 4 s T^{2} } \nu ^{2} ) \,
e^{-s M_{0} } \; \int \frac{ d^{3}p }{ (2\pi )^{3} } \; \tilde{ \eta }
(\vec{p}) \, \vec{p}^{2} \, \tilde{ \eta }(-\vec{p}) \; .
\label{eq:14}
\en
Gathering together all terms which contribute to the effective action
(\ref{eq:8}), and identifying $M_{0}$ and $\eta (\vec{x})$ with
$M(\vec{x})$ is tedious but straightforward. The final result is
\bea
\Gamma [M(\vec{x})] & = & \frac{1}{32 \pi ^{2} } \frac{1}{T} \int
d^{3}x \; \{ ( 1 + u f_{0}(u)) \frac{1}{12 M} \vec{\nabla } M \vec{\nabla } M
\; + \; U(M) \}
\label{eq:15} \\
U(M) &=& \frac{1}{2} M^{2} ( \ln \frac{M}{\mu ^{2} } -\frac{1}{2} )
\, - \, \frac{3}{2 \lambda _{R} } M^{2} \; - \;
M^{2} \, \frac{1}{u^{2}} f_{3}(u) \; ,
\label{eq:16}
\ena
where $u= M/4 T^{2}$ and the functions $f_{\epsilon } (x) $ are defined by
\be
f_\epsilon (x) \; = \; \sum_{\nu \not= 0} \int _{0}^{\infty }
\frac{ ds }{ s^\epsilon } \; e^{-sx } \; e^{- \frac{  \nu ^{2} }{s} } \; .
\label{eq:17}
\en
The functions $f_{\epsilon}$ decay exponentially for large
$x$~\cite{la3}, i.e.,
\be
f_{\epsilon }(x) \; = \; 2 \, x^{ \frac{ \epsilon -1 }{2} } \,
K_{\epsilon -1 }( 2 \sqrt{x} ) \; \approx \;
\sqrt{ \pi } \, x^{ \frac{ 2 \epsilon -3 }{4} } \, e^{ -2 \sqrt{x} } \; ,
\label{eq:22}
\en
where the $K_{\alpha }(x)$ are the modified Bessel functions of the second
kind. For $x \rightarrow 0$, the function $f_{3}$ approaches a
finite value whereas $f_{0}(x)$ diverges, i.e.,
\be
\lim _{x \to 0} f_{3}(x) \; = \; 2 \zeta (4) \; , \hbo
f_{0}(x) \sim \frac{\pi }{2}
\frac{ 1 }{ x^{3/2} } \; {\rm as}\; x\to0.
\label{eq:23}
\en
The renormalisation point dependence of the effective potential
(\ref{eq:16}) can be removed in favour of the zero-temperature value
of the scalar condensate~$M_c$, defined by the minimum of
(\ref{eq:16})~\cite{la3}, which gives
\be
U(M) \; = \; \frac{1}{2} M^{2} (\ln \frac{M}{M_{c}} - \frac{1}{2} )
\; - \; M^{2} \, \frac{1}{u^{2}} f_{3}(u) \; .
\label{eq:18}
\en
The effective potential $U(M)$ and the function $1+u f_{0}(u)$ are
shown in figure~2 for $T=0.25 \, \sqrt{ M_{c} }$ and it can be seen
that at this temperature, the false minimum of $U(M)$ at $M=0$ has
higher energy density than the true ground state.  The crucial
additional observation is that in this false-vacuum region ($u
\rightarrow 0$), the function $1+u f_{0}(u)$ becomes large. A large
prefactor in front of the kinetic term suppresses space dependent
fluctuations implying that tunneling from the false to the true ground
state is suppressed.

One might argue, on the other hand, that if the prefactor of the
kinetic term is large, the derivative expansion itself is invalid.
However, we are interested in configurations given by the classical
equation of motion, and so we need not calculate the effective action
for arbitrary field configurations. In fact, we will see that the
large prefactor enforces classical configurations with small gradients
and that higher derivative terms are suppressed even further.  In
particular, the kinetic energy for classical solutions is always small
compared to the effective potential, justifying the expansion for our
purposes.

\bigskip
{\it 3.\ The Slow-Rollover phase transition }

In this section we investigate the type of phase transition the
inflaton undergoes, if it tunnels from the false ground state to the
true vacuum. To proceed further, it is convenient to introduce the
order parameter
\be
\varphi (\vec{x}) \; := \; \sqrt{ M(\vec{x})} \; .
\label{eq:19}
\en
In terms of this order parameter the effective action (\ref{eq:15})
is~($u=\varphi^2/4T^2$)
\bea
\Gamma [\varphi ] & = & \frac{1}{24 \pi ^{2} } \frac{1}{T} \int
d^{3}x \; \{ ( 1 + u f_{0}(u)) \frac{1}{2} \vec{\nabla }
\varphi  \vec{\nabla } \varphi \; + \; V( \varphi ) \}
\label{eq:20} \\
V(\varphi) &=&  \frac{3}{4} U \bigl( M(\varphi ) \bigr) \; = \;
\frac{3}{8} \varphi^{4} ( \ln \frac{ \varphi ^{2} }{ M_{c} } -\frac{1}{2} )
\; - \; \varphi^{4} \, \frac{1}{u^{2}} f_{3}(u) \; .
\label{eq:21}
\ena

As can be seen in (\ref{eq:23}), the prefactor of the kinetic term
diverges when (classically) the inflaton is in the false ground state
at some finite temperature ($\varphi / T \rightarrow 0$). This would
imply that no tunneling could occur and the inflaton would never leave
the false vacuum.
Note, however, that if an infrared cutoff $m_{0}$ is used in the
definition of the function $f_{0}(x)$ in (\ref{eq:17})
\be
f_{0}(x) \; \rightarrow \; \sum_{\nu \not= 0} \; \int _{0}^{\infty }
ds \; e^{-s m_{0} } \; e^{-sx} e^{- \frac{ \nu ^{2} }{ s } } \; ,
\label{eq:21a}
\en
the prefactor of the kinetic term becomes finite. In the very early
universe, of course, there is a natural infrared cutoff given by the
horizon scale (i.e., $H^{-1}$ where $H=\dot R/R$ is the Hubble
parameter with $R$ the scale size of the universe). This cutoff should
be used in any action, though for non-divergent quantities it makes a
negligible difference at and below the GUT scale. We can very crudely
estimate the size of this prefactor (at $\varphi=0$) for a cut-off
$m_0\simeq H^2\simeq\pi M_c^2/2m_{Pl}^2$ where $m_{Pl}$ is the Planck
mass, if the inflationary transition occurs at the GUT scale~($T_c\sim
M_c^{1/2}\sim10^{15}~{\rm GeV}$). We have
\be
\frac{ m_{0} }{ T_{c}^{2} } \; \sim \; 8\pi {M_c\over m^2_{Pl}}\;\sim\;
10^{-8} \label{eq:21b}
\en
Using the asymptotic form for $f_0$ given by (\ref{eq:23}), the
prefactor of the kinetic term in the false vacuum state is then
\be
1 \, + \, \frac{ m_{0} }{ T_{c}^{2} } \, f_{0}
(\frac{ m_{0} }{ T_{c}^{2} } ) \; \approx \; \frac{ \pi }{2}
\frac{ T_{c} }{ \sqrt{ m_{0} } } \; \sim \; 10^4
\label{eq:21c}
\en

We note here in passing, that the calculation in the previous section
is for time-independent fields under the assumption of thermodynamic
equilibrium, and that neither of these assumptions can be exactly
satisfied in the early universe (in fact, for the Robertson-Walker
metric it is not even possible to define equilibrium phase space
distributions~\cite{be88}) though of course, we know they are very
good approximations for many epochs in the early universe. We expect
however, that the effect of the finite horizon size would be one of
the dominant violations of these assumptions at and above the GUT
scale (assuming no other phase transitions are occurring at that time).
A more quantitative estimate of the magnitude of the
prefactor~(\ref{eq:21c}) would require a more refined calculation.

Perhaps a more appropriate way of proceeding from this point would be
to calculate the bubble nucleation rate due to tunneling through the
temperature dependent barrier in (\ref{eq:21}) (assuming as usual,
that the processes of thermalisation and tunneling occur on two widely
different time scales), but suppressed by the kinetic-energy
prefactor. There are indications \cite{trans}, however, that the
transition may be only very weakly first order, or even second order,
implying that such a calculation would not appropriate at this level
of approximation and is thus beyond the scope of this letter.

The previous section does tell us, though, that in general not only
the effective potential, but also the prefactor of the kinetic term is
temperature dependent, and might be large for false vacuum
configurations. This fact is well known in the description of
collective phenomena in many-body systems \cite{rish}. In order to
further investigate the implications of the temperature dependent
prefactor of the kinetic term, we simulate its effect by using the
following effective action
\be
\Gamma [\varphi ] \; = \; \int d^{4}x \; \{ \frac{1}{2}
F( \frac{ \varphi }{T} ) \partial _{\mu } \varphi \partial _{\mu } \varphi
\; + \; V(\varphi ) \}\; ,
\label{eq:24}
\en
where
\bea
F(x) &=& 1 + c \, e^{-x},
\label{eq:25} \\
V(\varphi ) &=& \frac{3}{8} \varphi ^{4} \bigl( \ln
\frac{ \varphi ^{2} }{ M_{c} } - \frac{1}{2} \bigr)
\label{eq:26}
\ena
where $c$ is a (large) constant which, for now, we set equal to $10^{4}$
from~(\ref{eq:21c}) and we see that the prefactor $F(x)$ exponentially
approaches $1$ for large $x$ but becomes large for small $x$. We note
at this point, that (\ref{eq:24}) describes a phase transition and
Green's functions in a heat bath and that once inflation starts, and
the particle interaction rate becomes much less than $H$, this is is
no longer a good description \cite{cl92}. We therefore assume that at
some $T_s<T_c$ the temperature in~(\ref{eq:24}) becomes essentially
fixed and we take~(\ref{eq:24}) thereafter as a kinetic theory
description \cite{bo93}, though of course the effective
temperature of particle phase space distributions cools in the usual
way (i.e., $T\sim R^{-1}$ for massless particles and $T\sim R^{-2}$
for massive ones).

To study the transition of the inflaton from the false ground state to
the true ground state we must look for stationary points of the
effective action (\ref{eq:24}). This is most easily achieved by
mapping the order parameter $\varphi $ onto an auxiliary field $\psi $
defined by

\be
\psi := \int _{0}^{\varphi } \sqrt{ F( \frac{ \varphi ' }{T} ) }
\, d \varphi ' \; .
\label{eq:27}
\en
Note that this mapping is invertible since $\sqrt{F}$ is always
positive. In terms of this auxiliary field $\psi $ the action is
\be
\Gamma [\psi ] \; = \; \int d^{4}x \; \{
\frac{1}{2} \partial _{\mu } \psi \partial _{\mu } \psi
\; + \; V_{mod}(\psi ) \} \; ,
\label{eq:28}
\en
where the modified effective potential is
\be
V_{mod} (\psi ) \; := \; V \bigl( \varphi ( \psi ) \bigr) \; .
\label{eq:29}
\en
This is precisely what we require, the prefactor of the kinetic term
in (\ref{eq:24}) has been eliminated and so (\ref{eq:28}) therefore
describes the time-evolution of the field $\psi $ moving in a modified
potential $V_{mod}$ in the standard way, i.e., varying the Einstein
action for spatially homogeneous fields gives
\be
\ddot\psi+3H\dot\psi+V'_{mod}=0 \label{eq:29a}
\en
where again $H$ is the Hubble parameter. We stress that the
time-evolution of the $\psi $-field is directly transferred to the
time-evolution of the order parameter $\varphi $ since the mapping
(\ref{eq:27}) is time independent.  Figure 3 shows $V(\varphi )$ and
$V_{mod}(\psi )$ for the choice (\ref{eq:25}) and (\ref{eq:26}).
$V_{mod} (\psi )$ is very flat in the region of the false vacuum,
enabling $\ddot\psi$ to be ignored and, as further described in the
following section, a slow-rollover phase transition to occur.

\bigskip
{\it 4.\ Inflation}

We now examine more closely the details and consequences of the phase
transition in the above model as a mechanism for inflation.

For inflation we would like the scale of symmetry breaking to be of
the order of the GUT scale and so we imagine $M_c$ in the above to be
of order $(10^{15} {\rm GeV})^2$.  Far above the critical temperature,
the absolute minimum of the potential is at the origin, and the field
will be localised at $\varphi=0$ in the standard way (see figure~1).
There is no problem with this mechanism in the model we describe, in
contrast to the usual slow-roll scenario where the coupling is so tiny
that an initial thermal state is unlikely to occur
\cite{li83,cl92,muw85}. The large prefactor of the kinetic term, which
the $\varphi$-field incurs while it is near $\varphi=0$, also confers
the following two advantages during this epoch. First, it suppresses
spatial gradients, making a smooth patch of homogeneous
$\varphi$-field more probable. By the same token, it suppresses
fluctuations of the $\varphi$-field within this patch, so that as the
critical temperature is approached and then passed it is unlikely that
thermal fluctuations will be strong enough to push the field to the
new absolute minimum. The field will thus supercool in its false
vacuum state.

Eventually, when the temperature gets low enough, the field will
commence its slow roll to the absolute minimum (or alternatively, it
will tunnel through the remaining potential barrier and then start to
slow-roll). As we saw in the previous section, the slow-rolling is
again a consequence of the large kinetic energy prefactor. We now
examine this era in a little more detail.

For the phenomenological function $F$ we have chosen in eqn.
(\ref{eq:25}) it is in fact possible to perform the integral
(\ref{eq:27}) analytically. For $c\gg1$ and $\varphi/T\ll1$,
corresponding to large prefactor and hence slow rolling,
(\ref{eq:27})~becomes
\be
\psi\simeq2\sqrt{c}T\left(1-e^{-\varphi/2T}\right)\simeq\sqrt{c}
\varphi. \label{eq:30}
\en
Substituting this into (\ref{eq:29}) (and setting $T=0$ to remove
the $T$-dependent barrier as we are only concerned with slow-rolling)
gives
\be
V_{mod}(\psi)\simeq{3\psi^4\over8c^2}\left(\ln\left({\psi^2\over cM_c}
\right)-{1\over2}\right) \label{eq:31}
\en
during slow-rolling {\it only}. Of course (\ref{eq:30}) means that
(\ref{eq:31}) has the same Coleman-Weinberg form as (\ref{eq:26}),
however the extra factor $c^{-2}$ makes the effective ``coupling''
for the $\psi$-field extremely small because we expect $c$ to be very
large. It also makes analysis during slow-rolling particularly simple,
slow-rolling on the Coleman-Weinberg potential having been studied
many times before~\cite{ko90,cms89}.

In order to justify the derivative expansion for our purposes, we
estimate the magnitude of the kinetic term during the slow roll period.
Using the equation of motion for the auxiliary field $\psi $ (\ref{eq:29a})
(ignoring $\ddot \psi $ during slow rolling) and (\ref{eq:30}) we have
\be
c \; \dot \varphi^2 \; \approx \; \bigl(
\frac{ V'_{mod} }{ 3 H } \bigr) ^{2} \; = \;
O( \frac{1}{ c^{4} } ) \; ,\label{eq:31a}
\en
which is thus small compared to the intrinsic scale
(e.g.,\ the zero-temperature vacuum-energy density) of the effective
potential.

Ensuring that the slow-rolling does not conflict with bounds on
density fluctuations ($\delta\rho/\rho$) provided by the CMBR and
assuming such fluctuations are also responsible for large scale
structure ($\delta\rho/\rho\simeq10^{-5}$) allows one to test the
value for the parameter $c$ and also, as usual, implies inflation
lasts long enough to solve the horizon, flatness and monopole
problems. We first of all note that the energy density and pressure
of the $\varphi$-field have the following form
\bea
\rho_\varphi &=& {1\over2}F(\varphi/T)\dot\varphi^2+V(\varphi)
\label{eq:32} \\
p_\varphi &=& {1\over2}F(\varphi/T)\dot\varphi^2-V(\varphi)
\label{eq:33}
\ena
plus gradient terms proportional to $(\nabla\varphi)^2/R^2$ which we
ignore as usual. Now, even though the function $F$ is large for
$\varphi$ near zero, during slow rolling, the potential term will
still dominate from (\ref{eq:31a}). This implies that the amplitude of
density fluctuations generated by inflation is given by
\be
{\delta\rho\over\rho}\simeq {\delta\varphi V'(\varphi)\over
c\dot\varphi^2} \simeq {\delta\psi V'_{mod}(\psi)\over
\dot\psi^2} \label{eq:34}
\en
where we have used (\ref{eq:29}) and (\ref{eq:30}). The last
expression is just the usual form for fluctuations generated by a
weakly coupled field so that we expect $\delta\psi\simeq H/2\pi$ and
from the slow-rolling condition $3H\dot\psi\simeq-V'(\psi)$ (i.e.,
(\ref{eq:29a}) with negligible $\ddot\psi$), and where during inflation
$H^2\simeq const\simeq\pi M_c^2/2m^2_{Pl}$, from the $T=0$
$\varphi$-potential. So we get, again as expected,
\be
{\delta\rho\over\rho}\simeq {3 H^3\over V'_{mod}} \label{eq:35}
\en
Pursuing the standard analysis further we find that we can use
the approximation $V'_{mod}(\psi)\simeq \psi^3/c^2$ during the
slow roll and the value of $\psi$, $N$ $e$-folds before the
end of inflation is $\psi(N)\simeq(8/3c^2)^{1/2}\, N^{3/2}$.
Using these in (\ref{eq:35}) and solving for $c$ yields
\be
c\simeq\left({8N^3\over3}\right)^{1/2}\left({\delta\rho\over\rho}
\right)^{-1}. \label{eq:36}
\en
So, for generating cosmological structure, which requires $N\simeq50$
and $\delta\rho/\rho\simeq10^{-5}$ from COBE, we would need a value
$c\simeq10^7$. This is somewhat larger, but nevertheless encouragingly
similar to the crude estimate of (\ref{eq:21c}). Furthermore, if we
were to use, in (\ref{eq:21c}), the horizon scale and temperature
$T_s$, when inflation commences, rather than the values at the critical
temperature (where the minima in the effective potential are
degenerate), we would expect a larger value for $c$. We hope to verify
this in a more detailed calculation.

We make a remark here on slow rolling in our model. It might appear at
first from (\ref{eq:31}), that we have simply exchanged an unnaturally
small coupling constant for an unnaturally large multiplicative
constant. In the standard slow-rolling scenario, though, the tiny
coupling really is unexplained, excepting perhaps supergravity models
\cite{hol84} which, however, suffer from their own problems
\cite{ov83,jen86} (but incidentally, also offer a precedent for
considering non-minimal kinetic terms \cite{jen86}). In contrast,
however, in the model we consider here, the large value of $c$ simply
occurs from the dynamics of the theory --- in (\ref{eq:21c}) its
value is a consequence of the infrared cutoff given by the finite
horizon size.

After the slow rolling inflation transition, comes reheating, as the
$\varphi$-field drops into and oscillates about its new minimum. From
figure~2, it can be seen that at this true minimum the kinetic-term
prefactor has returned to its conventional value of ${1\over2}$, i.e.,
equations (\ref{eq:30}) and (\ref{eq:31}) are no longer good
approximations. The $\varphi$-field thus becomes a field with a
canonical kinetic energy, oscillating about its absolute
minimum --- perturbations in $\varphi$ are no longer suppressed and
most importantly, the fact that the coupling of $\varphi$ can have a
reasonable value means that the model does {\it not} of necessity have
a low reheat temperature, since other particles which may couple to
$\varphi$ can also have reasonable couplings.

In summary, we have described here a model which can undergo a
slow-rolling phase transition as a natural consequence of the dynamics
of the theory. The slow-rolling behaviour is due to a field dependent
prefactor of the kinetic term in the effective action, and its large
value in the false vacuum and during slow rolling is a natural
consequence of the non-perturbative dynamics of the theory. In
addition to natural slow-rolling, the model does not have some of the
usual problems associated with the conventional slow-rolling scenario.
In particular, there is no problem with thermal equilibrium as the
explanation for localising the field in the false vacuum, and there is
no necessity for a low reheat temperature.

\bigskip
\leftline{\bf Acknowledgements: }
\nobreak
K.L.\ would like to thank K.~Enquist for useful discussions concerning
the infrared cutoff in the early universe.
\medskip

\vfill \eject
\centerline{ \large Figure captions }

\vspace{2cm}
{\bf Figure 1:}
The temperature dependent effective potential $V(\varphi^2,T)$ for
different values of the temperature $T$. At high temperature $T\gg
T_c$ the minimum of potential is at $\varphi=0$. At zero temperature,
or the continuum (i.e., no periodicity in imaginary time), the
absolute minimum (corresponding to the non-trivial phase of
$\lambda\varphi^4$ theory) of the potential occurs at
$\lambda\varphi^2/6=M_c$, the magnitude of the scalar condensate.

\bigskip
{\bf Figure 2:}
The effective potential $U$ and the coefficient of the kinetic
term $1+u f_{0}(u)$ as functions of $M/M_c$ ($M\equiv\varphi^2$) at a
temperature $T=0.25 M_c^{1/2} $.

\bigskip
{\bf Figure 3:}
The potential of the order parameter $V(\varphi )$ and
the modified potential $V_{mod}(\psi )$ for $c=20$ in units of
$M_{c}^{2}$. Fields $\varphi $, $\psi $ in units of
$ M_c^{1/2}$.

%
%
%
%
%
%
%
%
%

%
%
\end{document}